\documentstyle[sprocl,epsfig,psfig]{article}

\def\be{\begin{equation}}
\def\ee{\end{equation}}
\def\bea{\begin{eqnarray}}
\def\eea{\end{eqnarray}}

\newcommand{\lsim}{\mathrel{\mathop{\kern 0pt \rlap
  {\raise.2ex\hbox{$<$}}}
  \lower.9ex\hbox{\kern-.190em $\sim$}}}
\newcommand{\gsim}{\mathrel{\mathop{\kern 0pt \rlap
  {\raise.2ex\hbox{$>$}}}
  \lower.9ex\hbox{\kern-.190em $\sim$}}}

\begin{document}

\title{FIRST RESULTS OF THE ROSEBUD DARK MATTER EXPERIMENT}

\author{}

\date{}

\maketitle

\begin{center}

S. Cebri\'{a}n$^{a}$,
N. Coron$^{b}$,
G. Dambier$^{b}$,
E. Garc\'{\i}a$^{a}$,
D. Gonz\'{a}lez$^{a}$,\\
I.G. Irastorza$^{a}$,
J. Leblanc$^{b}$,
P. de Marcillac$^{b}$,
A. Morales$^{a}$, \\
J. Morales$^{a}$,
A. Ortiz de Sol\'{o}rzano$^{a}$,
A. Peruzzi$^{a}$, \\
J. Puimed\'{o}n$^{a}$,
M.L. Sarsa$^{a}$,
S. Scopel$^{a}$,
J.A. Villar$^{a}$

\end{center}

\begin{center}
\begin{em}

$^{a}$Laboratorio de F\'{\i}sica Nuclear y Altas Energ\'{\i}as, \\
Universidad de Zaragoza,  50009 Zaragoza, Spain
\\
$^{b}$Institut d'Astrophysique Spatiale, B\^at. 121, \\
91405 Orsay Cedex, France

\end{em}
\end{center}


\abstract{ROSEBUD (Rare Objects SEarch with Bolometers UndergrounD)
is an experiment which attempts to detect low mass Weak Interacting
Massive Particles (WIMPs) through their elastic scattering off
Al and O nuclei. It consists of three small sapphire bolometers (of a
total mass of 100 g) with NTD-Ge sensors in a dilution refrigerator
operating at 20 mK in the Canfranc Underground Laboratory.
We report in this paper the results of several runs (of about 10
days each) with successively improved energy thresholds, and
the progressive background reduction obtained by improvement of the
radiopurity of the components and subsequent modifications in the
experimental assembly, including the addition of old lead shields.
Mid-term plans and perspectives of the experiment are also presented.}
\\
\\
PACS: 95.35.+d; 07.57.Kp; 07.62.+s
\\
{\it{Keywords:}} Dark matter; WIMPs; Underground detectors;
Bolometers

\section{Introduction}

Among the particle candidates to the non-baryonic dark matter of
the universe, the neutral, weak interacting massive particles
(WIMPs)---supposedly forming a significant part of the galactic
haloes---enjoy a prominent position. A distinguished WIMP is the
neutralino, the lightest stable particle of the minimal supersymmetric
extension of the Standard Model \cite{Gri}. Accelerator results settle a
lower bound of 20--30 GeV to the neutralino mass in most of the
models, but in unconstrained SUSY models masses as low as 2 GeV
could be allowed \cite{Gab96}. WIMPs of m $\sim 1-1000$ GeV with non-relativistic
velocity, can interact with the nuclei of a detector target producing
a nuclear recoil of a few keV, a fraction of which is visible
in the detector (depending on the nuclear target, the detector and
the mechanism of energy deposition). Because of the low interaction
rate and small energy deposition the WIMP detectors require very
low backgrounds and energy thresholds \cite{Mor3}. A further
contribution to the direct detection rate---in the very low
energy region $\leq 1$ keV--- could be provided by a non-thermal,
low velocity WIMP population recently proposed \cite{Dam98} and discussed
\cite{Bot00}.

Thermal detectors \cite{Cryo}, which measure the temperature increase produced
by the WIMP interaction in an absorber crystal, use more efficiently
the energy deposition of WIMPs than the conventional ionization detectors
because most of the nuclear recoil energy in WIMP scattering goes to heat.
They are true low energy detectors, where the visible energy is practically
the whole recoil energy (quenching factor close to one). Moreover, the
mechanisms and quanta involved in the physics of the detection imply
that they should have better energy threshold and energy resolution
than the conventional ionization detectors. In particular, the energy
resolution on nuclear recoils achieved by the sapphire bolometers of
the experiment reported in this paper is significantly better than that
obtained with HPGe diodes, like COSME \cite{Jmor92} (which features
a full width at half maximum (FWHM) energy resolution
$\Gamma (10 \rm keV)=0.4 \rm keV$).
In much the same way the energy threshold of the sapphire bolometers of
this work is also lower than that obtained in Ge-diodes looking for dark
matter \cite{Mor3,Jmor92,Reu91,Mor00}. Such low energy thresholds,
hopefully achieved by thermal detectors, make them valuable tools
in the search for phenomena \cite{Dam98} leaving very small energy in the
absorbers. So the cryogenic detectors are very well suited to explore
either low mass WIMPs or low velocity WIMPs \cite{Dam98}. They offer
moreover the possibility of using
different targets to tune up the sensitivity for different WIMP masses.
Finally, they can discriminate the background (electron recoils) from nuclear
recoils by collecting not only the phonon energy but also the charge (or light)
produced by the
ionizing component of the deposited energy (hybrid detection) \cite{Cha00,Gai00}.
The resulting background reduction permits to increase significantly the
sensitivity of WIMP searches allowing to explore regions below
$\sigma^{p} \sim 10^{-8}$ nbarn for medium mass ($\sim 50$ GeV) WIMPs---where
the neutralino configuration \cite{Bot98} relevant to interpret an annual
modulation effect \cite{Ber99} is located.
Although the radioactivity content of bolometers and their environment is still
higher than that of HPGe, significant progresses are being accomplished.

A WIMP search with small sapphire bolometers, ROSEBUD (Rare Objects Search with
Bolometers Underground), is being carried out in the Canfranc Underground
Laboratory (at 2450 m.w.e.) \cite{rosebud1,rosebud2,Ceb00} with the purpose
of exploring the WIMPs scattering off Al and O nuclei. It features three
small sapphire bolometers (with Ge thermistors) placed in
a small, mobile dilution refrigerator in the ultralow radioactivity environment of the
Canfranc underground site.

\section{Experimental set-up and performances}

\begin{figure}
\centerline{
\epsfxsize=12cm
\epsffile{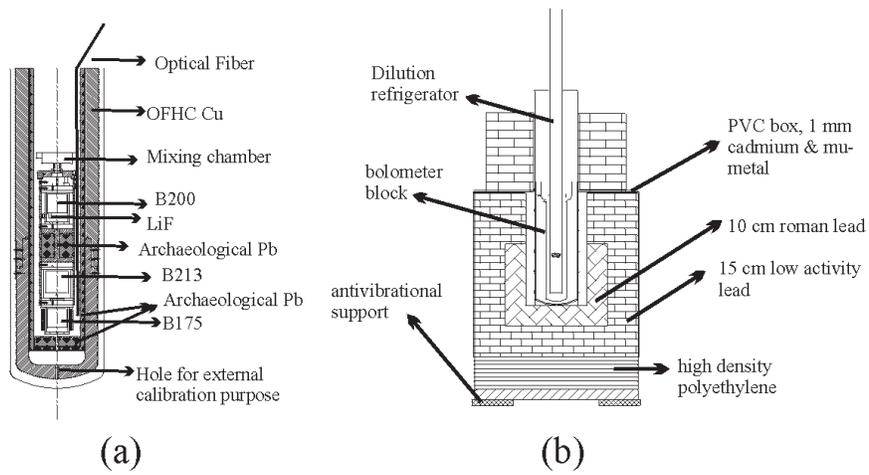}
}
\caption{{\bf (a)} Schematic view of the bolometer block in the first ROSEBUD run at Canfranc.
See text for explanation. {\bf (b)} View of ROSEBUD set-up.}\label{figure1}
\end{figure}

The first phase of the experiment consists of two 25g (B175 and B200)
and one 50g (B213)
selected sapphire bolometers operating inside a small
refrigerator (stick type, 6cm diameter, 100cm long) at 20 mK. One of the
25g sapphire crystals (B200), mounted on top of the set is part of a
composite bolometer (2g of LiF enriched at 96\% in $^6$Li are
glued to it) to monitor the thermal and fast neutron background of
the laboratory. A cylinder (34.5 mm diameter and 30.8 mm high) in Roman lead is placed
below B200 to screen the two lower sapphire absorbers (see Fig. \ref{figure1}) from
radioactivity in the components of the dilution unit.

The shielding of the ROSEBUD experimental box has an inner part inside
the dilution unit at helium bath temperature, consisting of a liner of
Roman lead, 4 mm thick, surrounding the bolometers (to shield
the intrinsic radioactivity of the 0.5 mm thick stainless steel walls of
the inner vacuum chamber) and of a 14.4 mm thick high purity copper (to shield
the radioactivity of the outer dewar). The ancient (2000 yr) Roman lead of the
shielding has very low contamination ($< 9$ mBq/kg on $^{210}$Pb and
$< 0.2$ mBq/kg on $^{238}$U and $< 0.3$ mBq/kg on $^{232}$Th). The external
part of the shielding (at room temperature) consists of 10 cm of Roman lead
bricks, 15 cm of low-activity lead (of about 100 yr old and of 30 Bq/kg of
$^{210}$Pb), then a 1-mm Cadmium foil and a mu-metal screen plus a plastic box
which close tightly the enceinte. A diagrammatic view of the experimental box within
the ROSEBUD dilution refrigerator and the shielding set-up are depicted in
Figure \ref{figure1} a) and b) respectively.
The experimental setup is installed within a Faraday cage in an acoustic
isolation cabin, supported by an antivibration platform, as required
to reach low thresholds. Slight modification of the experimental set-up
have been done along the successive runs for improving the background
and the energy threshold.

Power supply inside the cabin is provided by batteries and
data transmission from the cabin through convenient filters is
based on optical fibers. Pumps have vibration-decoupled
connections.

\begin{figure}
\centerline{
\epsfxsize=12cm
\epsffile{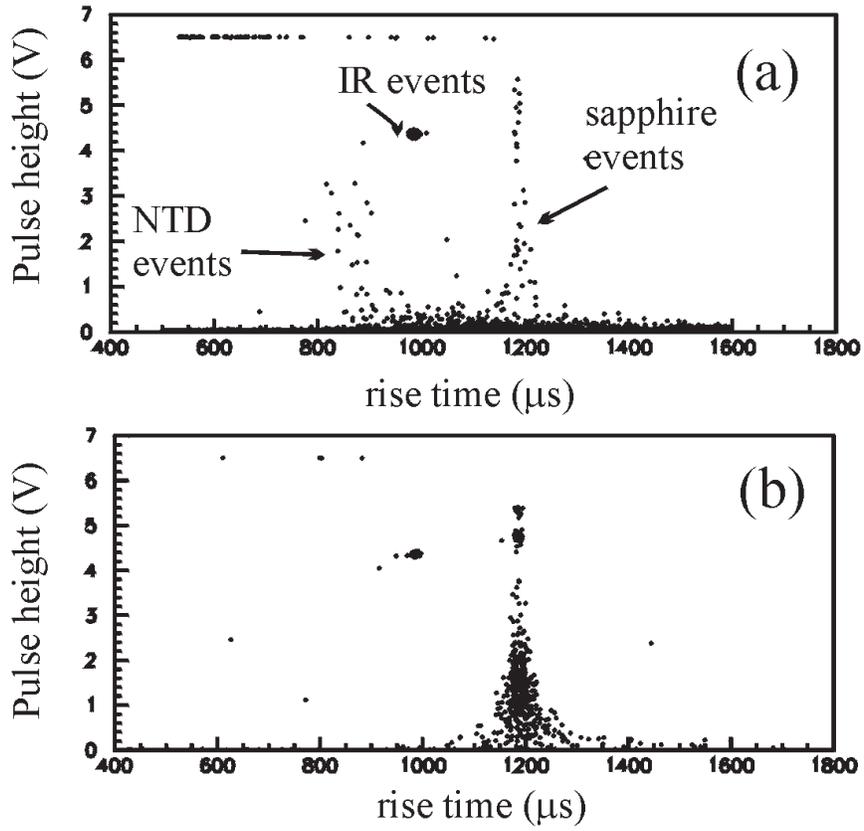}
}
 \caption{Rise time discrimination between IR pulses ($\sim 4$ V and
 $\sim 1000 \mu$s) and background events in the sapphire
 ($\sim 1200 \mu$s) recorded on bolometer B213 along a backgound measurement
 (a) and  a $^{57}$Co calibration (b). Events having low risetime in Fig.
 \protect \ref{fig_2}a can be attributed to background interactions
 in the NTD thermistor.}
  \label{fig_2}
  \end{figure}

Infrared (IR) pulses are periodically sent to the bolometers through
optical fibers in order to monitor the stability of the experiment.
The resulting heat pulses are useful to correct any sensitivity drift
of the bolometers during the experiment. The corresponding pulses are
quite similar in shape to the background or $^{57}$Co source events,
but slightly faster because they are mainly produced in
the thermistor, the sapphire being quite transparent to infrared
photons. This is illustrated in Fig. \ref{fig_2} where IR pulses
($\sim 4.4$ V) are clearly distinguished from background events in
the sapphire.

\begin{figure}
\centerline{
\epsfxsize=10cm
\epsffile{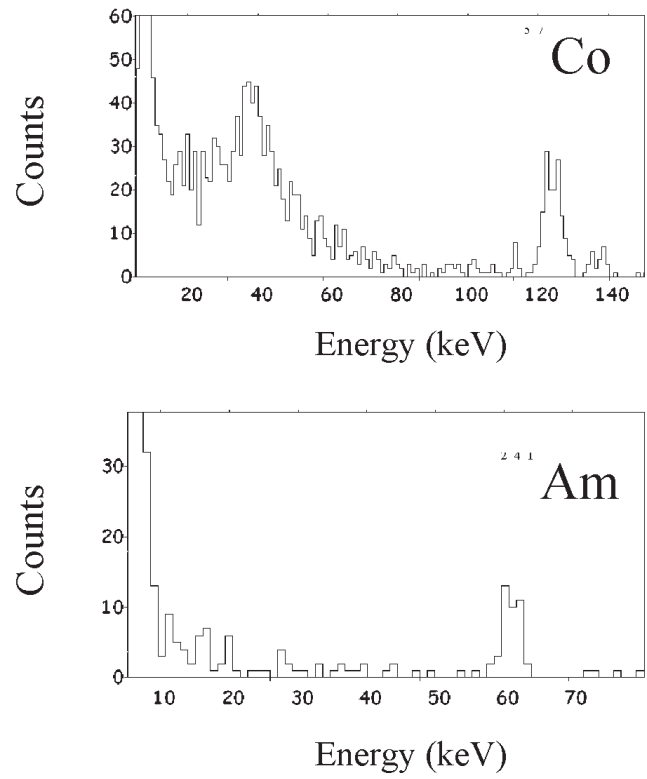}
}
 \caption{Typical calibration spectra from $^{57}$Co and $^{241}$Am sources
 in bolometer B175.}
  \label{figure2}
  \end{figure}

For absolute energy calibration of the two bolometers placed at the bottom
of the cryostat during the background tests, external $^{57}$Co and $^{241}$Am sources
are used (See Fig. \ref{figure2}).
The three lines (59.54, 122 and 136 keV) allow to extrapolate
(assuming linear response) the
calibration to lower energies. For the upper one (B200), an
internal $^{241}$Am source (of 3 Bq of activity) has been used. The
alpha emission (5.5 MeV) has been
degraded to 3 MeV by mylar foils in order to leave clean the region where the
neutron signal (above the Q-value of the $^6$Li(n,$\alpha$) reaction, 4.78 MeV)
should be found.

The bolometers are suspended inside an OFHC copper frame (coupled to the
mixing chamber of the dilution refrigerator) by Kevlar wires tensioned to
drive the mechanical resonant frequency to the highest possible value
(above 400 Hz). The refrigerator operating point was just below 20 mK
(mixing chamber temperature) and the bolometers were working at 28 mK
with a bias of 100-150mV on 40M$\Omega$ load resistors. The first
tests in Canfranc have shown that microphonic and electronic noise level is
quite good, about 2nV/Hz$^{1/2}$ at low frequency (30Hz) on bolometer B175.

\begin{figure}
\centerline{
\epsfxsize=8cm
\epsffile{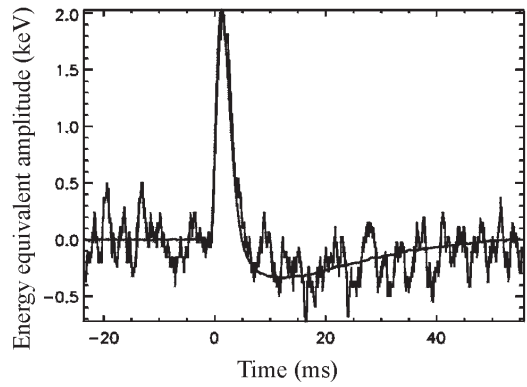}
}
\caption{A pulse of about 2 keV obtained at Canfranc with bolometer B213.
A normalized 100 keV event is overplotted.}\label{figure3}
\end{figure}

The bolometers were tested previously in Paris (IAS)
showing very good performances \cite{rosebud1}. (FWHM energy resolution
of $\Gamma (1.5 \rm keV) = 120 \rm eV$ and energy threshold around 300 eV).
Typical sensitivities
obtained (in Canfranc) are in the range of 0.3-1 $\mu$V/keV. Overall resolutions of
3.2 and 6.5 keV FWHM were typically obtained with B213 and B175, respectively, at 122
keV. With B175, 2.8 keV FWHM was obtained at 59.5 keV.
Typical pulses have decay and rise times between 2-4 ms and
600-1300 $\mu$s, respectively. A low energy background pulse in
bolometer B213 can be seen in Fig. \ref{figure3}. It corresponds to
an energy of about 2 keV.

Radiopurity measurements of some of the ROSEBUD internal components
(NTD sensors, sapphire crystals, copper frames...) were carried out at
the underground laboratories of Modane (LSM) and Canfranc using
ultralow background germanium detectors and the results reported
in Ref. \cite{rosebud2}. The substitution of the set-up components
with radioactive impurities by other cleaner components
would likely make them decrease down to levels acceptable for the
first phase of the experiment, as it has actually happened in
the runnings reported in the next paragraph. The upper limit on the
total background produded by those impurities was estimated \cite{rosebud2}
to be about 5 counts.keV$^{-1}$.kg$^{-1}$.day$^{-1}$ at 10 keV. This rate
is still one to two
orders of magnitude worse than that obtained in good germanium detectors
\cite{Mor3,Mor00,Bau99}.

\section{Results and prospects}

ROSEBUD has performed a series of short runnings in
different conditions (of a duration of a few days each) to
optimize the background, minimize microphonics and improve the
performances of the bolometers, before going to larger running
time.

\begin{figure}
\centerline{
\epsfxsize=10cm
\epsffile{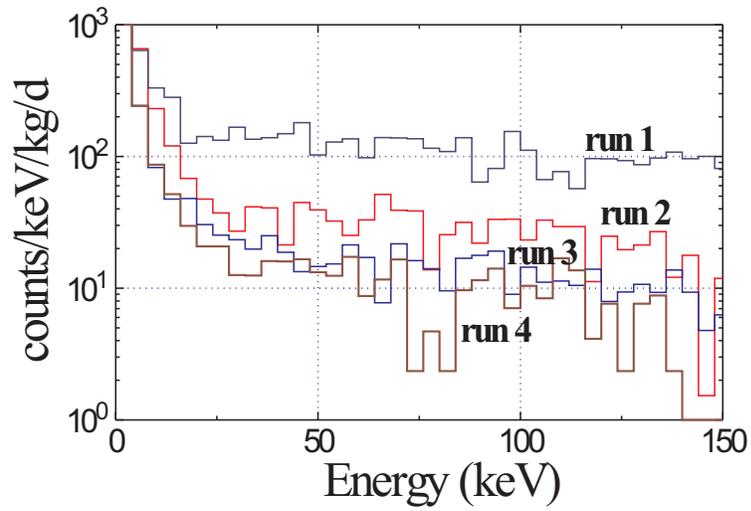}
}
\caption{The four successive background spectra of ROSEBUD obtained after
replacement of several components.}
\label{fig5}
\end{figure}

\begin{figure}
\centerline{
\epsfxsize=10cm
\epsffile{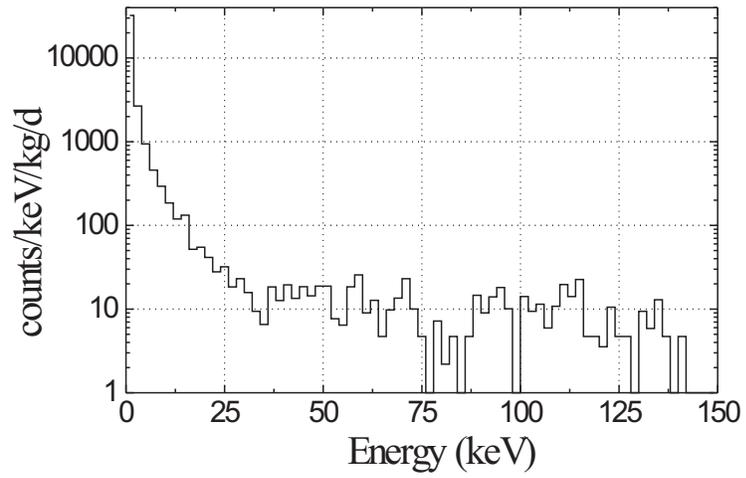}
}
\caption{Spectrum of the last running (Run 4) corresponding to 2.1 days
of exposure of bolometer B213 of 50 g. Conservatively, the cut in rise
time has been applied only down to 25 keV.}
\label{fig_7}
\end{figure}

\begin{figure}
\centerline{
\epsfxsize=10cm
\epsffile{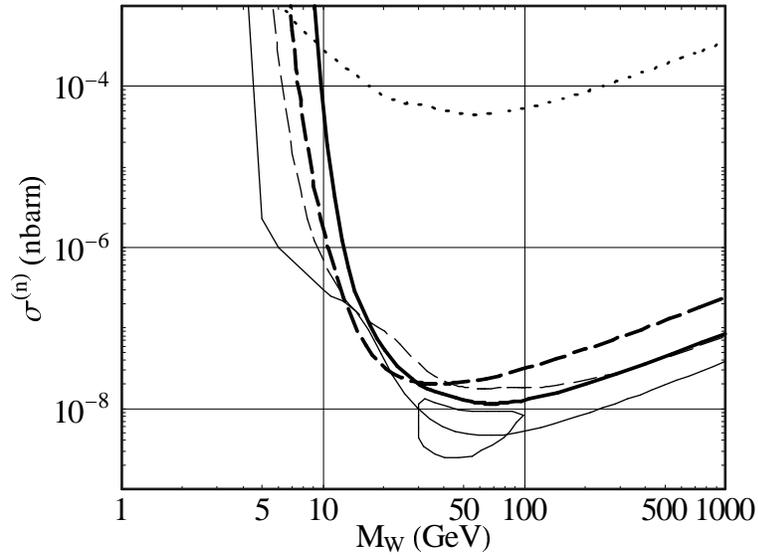}
}
\caption{Exclusion plot (dotted line) cross-section versus WIMP mass for
spin-independent WIMP-nucleon interaction at the 90\% C.L.
derived from the spectrum of Figure \protect \ref{fig_7}.
Other exclusion limits obtained with Ge and NaI detectors are also
shown: The Ge combined bound from published data (thin dashed line) and
the more recent COSME-2 (thick dashed line) and IGEX (thick solid line)
results are shown together with the DAMA (NaI-0) exclusion contour
(thin solid line) and the DAMA (NaI-1,-2,-3,-4) annual modulation region
(closed contour).}\label{fig_8}
\end{figure}

In a first testing measurement in the shielding configuration described
above a background as large as 120 counts/keV/kg/day around 100 keV was
obtained (Run 1). To understand and reduce this background, several components of
the cryostat were measured in the ultralow background HPGe testing bench at
Canfranc. After subsequent modifications in the cryostat, like the
replacement of the old multilayer glass fiber insulation and the reduction
of the charcoal mass, a background level of about 30 counts/keV/kg/day around
100 keV was obtained in bolometers B175 and B213 (Run 2).
Further reduction of the internal background was achieved by replacing some
other internal pieces of the cryostat and removing completely the coconut
charcoal, which was replaced by artificial carbon tissue (Actitex), and adding a
further inner lead shielding (15 cm high, 15 cm external diameter and 5 mm
thick) \cite{Ceb00}. The resulting
background was 15 counts/(keV kg day) around 100 keV (Run 3).
The next efforts were focused in eliminating possible
radon contaminations inside the nitrogen reservoir of the cryostat
as well as in changes in the inner configuration, in particular, removing
the bottom bolometer B175 and putting instead an additional archaeological
lead cylinder 4 cm thick. A short running to test the progress of this
reduction shows a background of $\sim 18$ counts/(keV kg day) in the region
from 25 to 50 keV, of $\sim 11$ counts/(keV kg day) between 50 and 100 keV and
$\sim 7$ counts/(keV kg day) between 100 and 150 keV (Run 4).
The four successive background spectra obtained in the corresponding set-up
configurations enumerated above are shown in Fig. \ref{fig5}. A rise time
cut has been applied to eliminate background interactions in the NTD thermistor
(see Fig. \ref{fig_2}).
By taking the spectrum of Run 4 (see Fig. \ref{fig_7})
---corresponding to an exposure of 2.1 days
with bolometer B213 of 50 g of Al$_2$O$_3$--- obtained with a
conservative rise time cut ---applied only down to 25 keV---
a 90\% C.L. exclusion plot
$\sigma$(m) has been derived for spin-independent interaction with
the conservative assumption of a 2-keV low energy resolution and a
2-keV threshold. The exclusion plots have been
derived by requiring the predicted WIMP signal in the 2-keV energy bin
from 2 to 100 keV to be less than or equal to the 90\% C.L.
upper limit of the (Poisson) background counts recorded there.
The derivation of the interaction rate signal supposes that
the WIMPs form an isotropic, isothermal, non-rotating halo of density
$\rho = 0.3$~GeV/cm$^{3}$, have a Maxwellian velocity distribution
with $\rm v_{\rm rms}=270$~km/s (with an upper cut corresponding to
an escape velocity of 650~km/s), and have a relative Earth-halo velocity
of $\rm v_{\rm r}=230$~km/s. The quenching factor (nuclear versus electron
recoil efficiency) has been taken unity ($0.98 \pm 0.05$ for sapphire
bolometers measured at 100 keV recoil energy \cite{Zho94}). A Fermi
nuclear form factor has been used.
Fig. \ref{fig_8} shows the contour line of this exclusion compared with
that obtained with Ge (Ge-combined from published data \cite{Jmor92,Reu91,Bau99,Ger3}
and the more recent COSME-2 \cite{Mor3,NJP00} and IGEX
\cite{Mor3,Mor00,NJP00}) and NaI (DAMA-0) detectors \cite{Ber96}.
The region singled out by the DAMA modulation effect is also shown \cite{Ber99}.

\begin{figure}
\centerline{
\epsfxsize=10cm
\epsffile{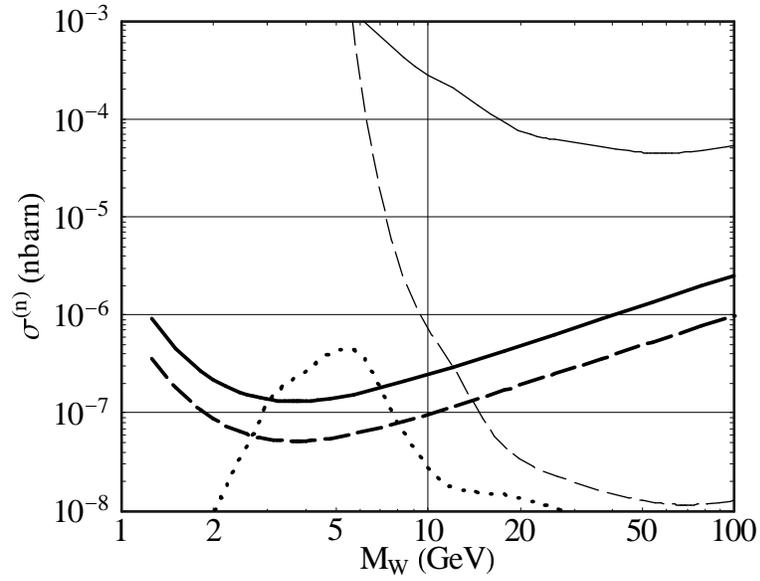}
}
\caption{Projected $\sigma$(m) exlcusion plots of the ROSEBUD
experiment for spin-independent coupling (assuming a flat
background of 1 c/keV kg day, a threshold of 300 eV and an energy
resolution of 100 eV), for the cases: 50 g of sapphire
operating 30 days (thick solid line) and along one year (thick dashed
line) compared with the unconstrained MSSM allowed region (dotted line). The
current ROSEBUD result (thin solid line) and the total Ge-combined bound
(thin dashed line) are also shown.}
  \label{fig6}
  \end{figure}

Fully understanding and significantly improving the background is the current
objective of ROSEBUD. With such purpose, measurements of the
radiopurity of individual components proceed and their removal done
when needed.
Assuming a flat background of 1 count keV$^{-1}$ kg$^{-1}$ day$^{-1}$, a
threshold energy of 300 eV and a FWHM energy resolution of 100 eV in the low
energy region (nominal values obtained for the bolometer B213 in the
Paris tests \cite{Mor3}), a projection of the ROSEBUD sensitivity is
shown in Fig. \ref{fig6} in terms of the WIMP-nucleon spin-independent
interaction exclusion plot.
To derive such plot, the chosen energy bin has been from 300 to 800 eV.
Projected exclusion contours $\sigma$(m) are
shown for two typical cases: 50 g of Al$_2$O$_3$ operating 30 days and
operating along one year. The area enclosed by the dotted line represents
the region for the
neutralino-nucleon cross section versus mass allowed by an unconstrained Minimal
Supersymmetric Standard Model (MSSM) where usual GUT relations among
gaugino masses are relaxed \cite{Gab96}. The exclusion plots
have been renormalized to the neutralino-nucleon cross section and,
for comparison, the most recent \cite{Mor00} Ge combined bound is also shown.
Assuming the mentioned projected performances, ROSEBUD will allow to explore
neutralino configurations corresponding to masses well below 10 GeV.

The second phase of the ROSEBUD program will deal with three different
bolometers: the current B213 sapphire bolometer of 50 g, a new 50 g
sapphire with cleaner components, and a germanium bolometer of 67 g,
operating together to investigate the target dependence of the WIMP rate,
provided that a reliable understanding of their background peculiarities
be achieved. As a further step in this line of research, the use of two
medium-size bolometers, one of 200 g of Al$_2$O$_3$ and the other a double
luminiscent bolometer (BGO or CaWO$_4$) is envisaged. Comparison of
the spectra of these different mass number bolometers will help to understand
the background and hopefully explore (within the same provisions) the
mass number dependence of the signal rate as a genuine characteristic of
WIMPs.

\section*{Acknowledgements}
This work has been supported by the Spanish CICYT (contract
AEN99-1033), the French CNRS/INSU (MANOLIA program) and by the
EU Network Contract ERB-FMRX-CT-98-0167. We greatly acknowledge
the help of J.P. Moalic, P. Pari and O. Testard for the cryogenics
design. We are also endebted to C.Goldbach and G. Nollez for their
contribution to the radioassays of various materials in the Modane
Underground Laboratory (Frejus), as well as for fruitful discussions.
We also thank J.L. Picolo, P. Cassette (LNHB) and P. Bouisset
(LMRE) for their radioactive measurements of some inner
components. N.Coron, G.Dambier and P. de Marcillac acknowledge
financial support of the DGA-CAI Europa Program.

\newpage

\newpage

\newpage

\newpage

\newpage

\newpage

\newpage

\end{document}